\documentclass[11pt]{article}
\usepackage{epsf,latexsym,times}

\begin{document}

\title{ Elastic properties of model porous ceramics
\footnote{Submitted to the {\em Journal of the American Ceramic Society}}}

\author{ {\normalsize Anthony P. Roberts$^{\ddagger,\dagger}$ and
Edward J. Garboczi$^{\ddagger}$}
\\
{\normalsize $^\ddagger$Building Materials Division,} \\
{\normalsize National Institute of Standards and Technology, 
Gaithersburg, MD 20899, USA }
\\
{\normalsize $^\dagger$Centre for Microscopy and Microanalysis,} \\
{\normalsize University of Queensland, St.\ Lucia, Queensland 4072,
Australia}
\\ \\
}

\date{ {\normalsize June 20, 2000}}

\maketitle

\vspace{-5mm}

\begin{abstract}
The finite element method (FEM) is used to study the influence
of porosity and pore shape on the elastic properties of
model porous ceramics.
The Young's modulus of each model was found to be
practically independent of the solid Poisson's ratio. At a 
sufficiently high porosity,
the Poisson's ratio of the porous models converged to a fixed value
independent of the solid Poisson's ratio. The Young's modulus of
the models is in good agreement with experimental data. We provide
simple formulae which can be used to predict the elastic properties
of ceramics, and allow the accurate interpretation of empirical
property-porosity relations in terms of pore shape and structure.
\end{abstract}

\section{Introduction}

\noindent
The elastic properties of two-phase (solid-pore) porous materials 
depend on the geometrical nature of the pore space and solid phase,
as well as the value of
porosity~\cite{Coble_K56,DeanLopez83,Rice_Eval_MSA,HerBax99}.
Relevant aspects
of porous materials may include pore shape and size
as well as the size and type of the interconnections
between solid regions. These features, which generally lack
precise definition, comprise the microstructure of
the material.  In order to predict properties, or properly
interpret experimental property-porosity relationships,
it is necessary to have an accurate method of
relating elastic properties to porosity and microstructure.
In this paper we use the finite element method to derive
simple formulae that relate Young's modulus and Poisson's
ratio to porosity and microstructure, for three different models of
microstructure.

There have been several different approaches to deriving
property-porosity relations for porous materials. Formulae
derived using the {\em micro-mechanics}
method~\cite{Hashin83,AboudiBook,ChristensenBook}
are essentially
various methods of approximately extending 
exact results for small fractions of spherical or ellipsoidal pores
to higher porosities. This includes the
differential~\cite{McLaughlinDM77} and self consistent
methods~\cite{HillSCM,BudianskySCM,Wu66,Berryman80_ell}
as well as the commonly used semi-empirical
correction to the dilute result made by
Coble and Kingery~\cite{Coble_K56} to explain 
the properties of porous alumina. A drawback of this approach
is that the microstructure corresponding to a particular
formula is not precisely known; hence agreement or disagreement
with data can neither confirm nor reject a particular model.
A second problem is that these types of models provide no predictions
for the case where the microstructure is comprised of incompletely
sintered grains, which is a common morphology in porous ceramics.
A second class of results~\cite{Rice_Eval_MSA,Rice_Comp_MSA}
have been termed {\em minimum solid area} (MSA)
models. In this approach purely geometrical reasoning
is used to predict the elastic moduli based on the weakest
points within the structure. 
Again, the microstructure
that corresponds to the MSA predictions is not exactly known.
A number of semi-empirical relations have also been
proposed~\cite{DeanLopez83}, which generally
provide a reasonable means of describing data,
extrapolating results and comparing data among materials.
However, lacking a rigorous connection with microstructure, these results
do not offer either predictive or interpretive power.
Theoretical bounds~\cite{Hashin83,TorqRev91}
exist for the elastic properties, but the vanishing of the lower bound
for porous materials lessens their predictive power when the
upper bound does not provide a good estimate.
There are numerous other approaches, including the generalized
method of cells~\cite{AboudiBook,HerBax99}.

Another approach is to computationally solve the equations of elasticity for
digital models of microstructure~\cite{Garboczi95a,Adlerelas1}.
In principle this can be done exactly.
However, large statistical variations and insufficient resolution,
have limited the accuracy of results obtained to date.
Only recently have computers
been able to handle the large three-dimensional models and number
of computations needed to obtain reasonable results. As input to
the method, we employ three different microstructural models that
broadly cover the types of morphology observed in porous ceramics.
The models are based on randomly placed spherical pores, solid spheres,
and ellipsoidal pores~\cite{TorqRev91}.
The centers of the pores or solid particles are
un-correlated which leads to realistic microstructures in which
both the pore and solid phase are interconnected.
The results, which can be expressed simply by two (or sometimes three)
parameter relations, correspond to a particular microstructure and
explicitly show how the properties depend on the nature of the porosity.
Therefore, the results can be used as a predictive tool for cases where the
microstructure of the ceramic is similar to one of the models,
and as an interpretive tool if the microstructure is unknown.
The numerically exact FEM results are compared with various well-known
micro-mechanics and MSA
results to determine how close an approximation a particular
formula provides for each model. In the FEM, we can freely vary 
the properties
of the solid phase, allowing us to determine the dependence of 
Young's modulus and Poisson's ratio on the solid Poisson's ratio as well 
as on the porosity.
This question has attracted recent interest in the ceramics
literature~\cite{RamAru93,Bocc94,Rice_Comm_PR}.

\begin{center}\noindent{\bf II. Computational Results}\end{center}

\noindent
A microstructure made up of a digital image is already naturally
discretized and so lends
itself to numerical computation of many quantities.
The finite element method uses a variational formulation of the linear elastic
equations, and finds the solution by minimizing the elastic energy via a fast
conjugate gradient method. 
The digital image is assumed to have periodic boundary conditions.
Details of the theory and programs used are
reported in the papers of Garboczi \& Day~\cite{Garboczi95a}
and Garboczi~\cite{Eds_manual}. 

In order to obtain accurate results using the FEM on models of random
porous materials,
it is absolutely necessary to estimate and minimize three sources of error:
finite size effects,
discretization errors, and statistical fluctuations. This has generally 
not been done in the past,
owing to limitations in computer memory and speed.  FEM results for random
microstructures do not have much meaning without such an error analysis.

The various sources of error are defined in the following way. 
First, the length scale of the microstructure is
fixed, usually by fixing the size of a typical pore (e.g.\, the spherical pore
radius).  The size of the system is then controlled by 
the side length of the cubic sample, denoted $T$. The size of $T$ compared to
the pore size controls how many pores will appear in the computational cell.
A real material has many thousands or more such pores. Errors can occur in
using a smaller number in a periodic cell to simulate a much larger number.
We vary $T$ in order to map out this effect.

Once a value of $T$ is chosen that minimizes finite size errors but is still
computationally possible, we next must consider the discretization error,
which comes about because we are using discrete pixels
to represent continuum objects. The number of pixels along each edge of 
the cubical unit cell is $M$, giving
a resolution of $dx$=$T/M$ (in units of $\mu$m per pixel, if $T$ is in $\mu$m). 
For the chosen value of $T$, a value of $M$ is chosen that also gives acceptable
discretization errors, usually on the order of a few percent.

Finally, when computing the properties of random materials, either 
computationally or experimentally, 
one must carefully choose the number of samples ($N_s$) over 
which the results need
to be averaged to produce acceptable uncertainties.
This value is again chosen, within computational constraints, to keep 
statistical fluctuations within a few percent.

\begin{figure}[b]
\centering \epsfxsize=14.5cm\epsfbox{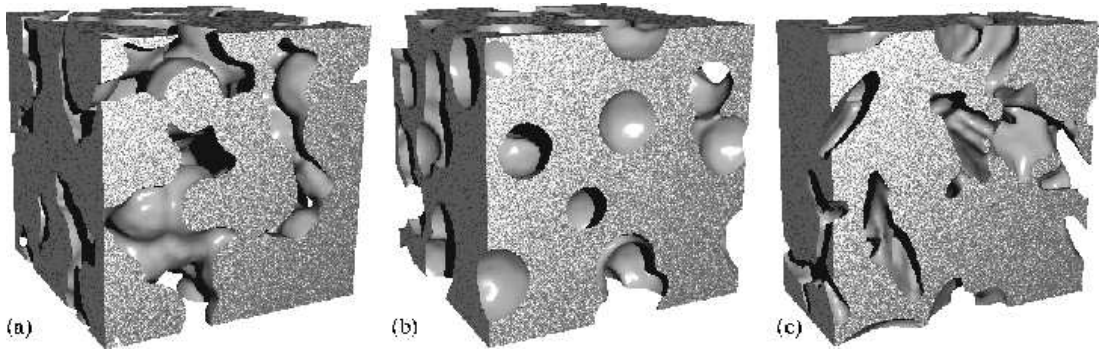}
\caption{ Showing pieces of the various models studied:  (a) overlapping solid
spheres, (b) overlapping spherical pores, and (c)
overlapping ellipsoidal pores.}
\label{gath3D}
\end{figure}

\noindent
{\bf Overlapping Solid Spheres.} 
Realizations of the overlapping solid sphere
model~\cite{TorqRev91,Weissberg63} are generated by
placing solid spheres at random points in the unit cell. This produces
a set of overlapping grains that mimic the microstructure of sintered
ceramic composites (see Fig.~\ref{gath3D}a). 
The space outside the solid grains 
is the pore space, with porosity $\phi$.
The pore phase is macroscopically connected above porosities of
$\phi\approx 0.03$ and the solid phase remains connected for values of
$\phi$ below $\approx 0.70$~\cite{TorqRev91}. Above $\phi$=0.7, the
solid phase is composed of isolated solid particles.
So between
$\phi = 0.03$ and $\phi = 0.70$, the 
overlapping solid sphere model is bi-continuous.
In ceramics the porosity is generally less then $0.40$, in this bi-continuous
regime.
We therefore consider the elastic
properties for $0.1 \leq \phi \leq 0.50$,
where the solid Poisson's ratio, $\nu_s$, varied over
the range $0.1 \leq \nu_s \leq 0.4$. 

To generate the microstructure we chose solid spheres of
radius $r$=1 $\mu$m.  Note that the elastic properties
are length scale invariant so the results apply to spheres
of any radius for which the continuum assumption holds.
A preliminary study showed that finite size errors were acceptably small
for cubic samples with edge length $T$=12 $\mu$m.
To study the discretization errors we generated one
realization of the model with porosity $\phi$=0.5
at $M$= 48$\dots$128.
The elastic properties depend quite strongly on resolution.
We found that the variation of Young's modulus with $M$ could be described
by the relation~\cite{Roberts95a} 
\begin{equation} E_{\rm{FEM}}(M)\approx E_0+a M^{-1}
\label{discEans}
\end{equation}
where $E_0$ can be identified as the continuum value (corresponding to
infinitely large $M$).  The same is true for Poisson's
ratio.
Even at $M=128$ the finite element
code overestimates the `exact' result for the Young's modulus by 30 \%.
Therefore, for the overlapping sphere model it is necessary to measure
the elastic moduli at three different values of $M$ and extrapolate the
results to $M\to\infty$.
We chose $N_s$=5 samples at each resolution and porosity,
except at $\phi$=0.5 where large statistical variations implied a
larger number of samples was necessary ($N_s$=10). Thus 30 different
realizations of the models were considered, each at 3 different
discretizations, for a total of $90$ models.

The statistical variation in Young's modulus and Poisson's ratio
for the case $\nu_s$=0.2 are shown in Table~\ref{props_summ}.
The error bars shown in the table are equal to twice the standard error
(S.E.=$\sigma/\sqrt{N_s}$ with $\sigma$ the standard deviation).
Therefore there is a 95\% chance that the ``true'' result lies
between the indicated error bars.
The results are accurate to within 20\% at $\phi$=0.5;
the error decreasing with porosity
to less than 10\% for $\phi \leq 0.30$. The expected
Gaussian distribution of the measured averages implies
that the results are actually more accurate than this.
For example, the anticipated relative errors are halved 
if a 68\% likelihood threshold is
used (i.e.,\ $\pm$ one standard error).

\begin{table}[t]
\caption{Elastic properties of the three models ($\nu_s$=0.2).
\label{props_summ}}
\begin{center}
\begin{tabular}{|c|c|c|c|c|c|c|}
\hline
\multicolumn{1}{|c|}{} &
\multicolumn{2}{|c|}{Overlapping solid spheres} &
\multicolumn{2}{|c|}{Overlapping spherical pores} &
\multicolumn{2}{|c|}{Overlapping ellipsoidal pores} \\
\hline
$\phi$ & $E/E_s$  & $\nu$ & $E/E_s$  & $\nu$ & $E/E_s$  & $\nu$ \\
\hline
0.1& 0.71   $\pm$   1\%   &     0.19     $\pm$      1\% &
     0.80   $\pm$     1\%  &      0.20   $\pm$     1\%&
     0.73   $\pm$      2\%   &   0.19  $\pm$          3\%\\
0.2& 0.47   $\pm$   2\%   &     0.18      $\pm$      4\% &
     0.62   $\pm$     2\%  &      0.20   $\pm$     2\%&
     0.52   $\pm$      3\%   &   0.18  $\pm$          4\%\\
0.3& 0.25   $\pm$   6\%   &     0.17     $\pm$      9\% &
     0.46   $\pm$     3\%  &      0.21   $\pm$     3\%&
     0.34   $\pm$      4\%   &   0.18  $\pm$          6\%\\
0.4& 0.12   $\pm$   13\%  &     0.15     $\pm$     25\% &
     0.33   $\pm$     4\%  &      0.21   $\pm$     4\%&
     0.20   $\pm$      3\%   &   0.18  $\pm$          4\%\\
0.5&0.039   $\pm$   22\%  &     0.15     $\pm$     21\% &
     0.21   $\pm$     8\%  &      0.22   $\pm$     9\%&
     0.11   $\pm$      4\%   &   0.18  $\pm$          6\%\\
\hline
\end{tabular}
\end{center}
\end{table}

\begin{figure}[t]
\centering \epsfxsize=8.3cm\epsfbox{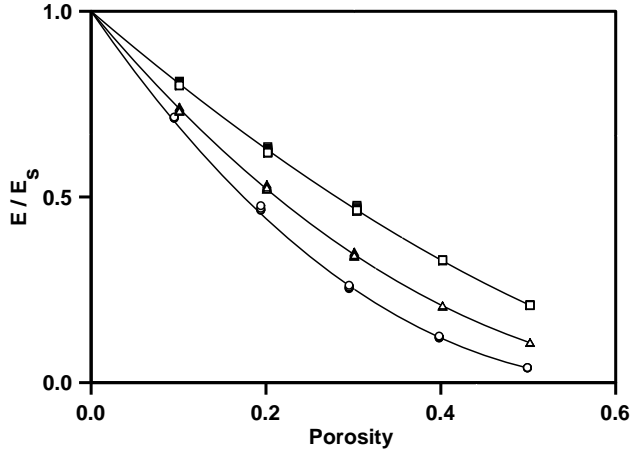}
\caption{
The Young's modulus of the three microstructure models.
The solid lines are empirical fits to the
equation $E/E_s=(1-\phi/\phi_0)^n$. Data is shown for 
overlapping solid spheres $(\circ, n=2.23, \phi_0=0.652)$,
spherical pores $(\Box, n=1.65, \phi_0=0.818)$ and ellipsoidal
pores $(\triangle, n=2.25, \phi_0=0.798)$ for $\nu_s=-0.1,\dots,0.4$.
Note that the $E$ is practically
independent of the solid Poisson's ratio in each case
[the different values of $E(\nu_s)$
at each porosity are almost indistinguishable].
\label{all_Evp}
}
\end{figure}

In addition to the above results we also computed the elastic
moduli of the 90 model microstructures at solid Poisson's ratios
$\nu_s$=0.1, 0.3 and 0.4. The statistical variation was
not significantly different from the case $\nu_s$=0.2.
Combined with the data for $\nu_s$=0.2
this covers most commonly occuring solids. The scaled Young's modulus
for each value of $\nu_s$ is plotted against porosity in
Fig.~\ref{all_Evp}. Remarkably, the scaled Young's modulus of the porous
material appears to be practically independent of $\nu_s$. 
This result has been proven to be exact
in 2-D~\cite{Garboczi_Circ_Holes92,Cherk92} and
appears to hold to a very good approximation in 3-D.
We found
that the Young's modulus data are well described by an equation of the form 
\begin{equation}
\frac{E}{E_s}=\left( 1- \frac \phi\phi_0 \right)^n 
\label{formE} \end{equation}
with $n$=2.23 and $\phi_0$=0.652 and $0 \leq \phi \leq 0.5$. Note that
$n$ and $\phi_0$ are empirical correlation parameters and
should not be interpreted as the percolation exponent and threshold,
respectively. Percolation concepts are generally valid closer
to the threshold $\phi_c\approx0.7$ (for this model) and a higher value of
$n$ is expected.  The computational cost of
accurately measuring the elastic properties increases
greatly as the percolation threshold is approached.

\begin{figure}[ht!]
\centering \epsfxsize=8.3cm\epsfbox{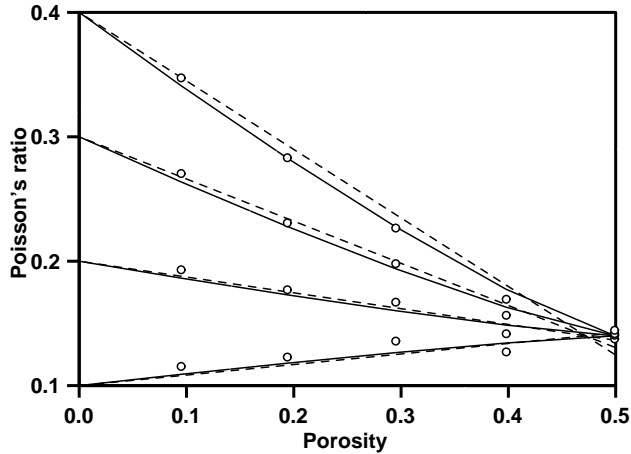}
\caption{
The Poisson's ratio of the overlapping solid sphere model as a function
of porosity for $\nu_s$=0.1--0.4.
The dashed lines are an empirical fit to
Eq.~(\ref{lineSform}).
The solid lines  correspond
to the three parameter relation given in Eq.~(\ref{percSform}), with the value
of all parameters given in the text.
The intercepts of the lines at zero porosity correspond
to the solid Poisson's ratio.
\label{oss_SvP}
}
\end{figure}

The Poisson's ratio of the porous material is shown in
Fig.~\ref{oss_SvP} as a function of $\phi$ and $\nu_s$, and appears
to be a flow diagram~\cite{Garboczi_Circ_Holes92},
where the Poisson's ratio asymptotically
approaches a fixed point, independently of the value of the solid
Poisson's ratio.  This flow diagram has been analytically 
proven to hold in 2-D,
when a percolation threshold exists at which the Young's modulus goes to
zero~\cite{Garboczi_Circ_Holes92,Cherk92}.
This flow diagram also appears to be valid in 3-D as 
well, within numerical uncertainty.
The Poisson's ratio data shown in Fig.~\ref{oss_SvP} 
can be roughly described by the simple linear relation,
\begin{equation}
\nu =\nu_s+\frac \phi\phi_0 (\nu_0-\nu_s) =
\nu_0 + \left(1- \frac \phi\phi_0\right)(\nu_s-\nu_0) 
\label{lineSform}
\end{equation}
with two fitting parameters $\nu_0$=0.140 and $\phi_0$=0.472.
A more accurate fit is obtained with the three parameter relation,
\begin{equation}
\nu= \nu_0 + \left(1- \frac \phi\phi_0\right)^m(\nu_s-\nu_0).
\label{percSform}
\end{equation}
with $\nu_0$=0.140, $\phi_0$=0.500 and $m$=1.22.

\begin{figure}[t]
\centering \epsfxsize=8.3cm\epsfbox{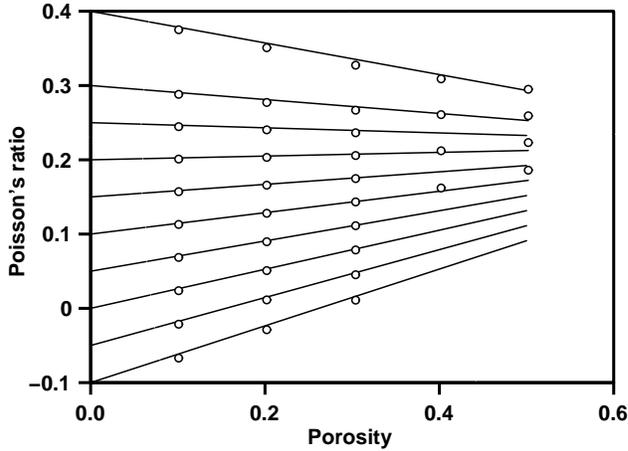}
\caption{Poisson's ratio of the overlapping spherical pore model as a
function of solid Poisson's ratio and porosity.
The lines are an empirical fit to
the relation $\nu=\nu_s+\phi/\phi_0 \times (\nu_0 -\nu_s)$ with
$\nu_0$=0.221 and $\phi_0$=0.840.
The intercepts of the lines at zero porosity 
correspond to the solid Poisson's ratio.
\label{E111_SvP}
}
\end{figure}

\begin{figure}[ht!]
\centering \epsfxsize=8.3cm\epsfbox{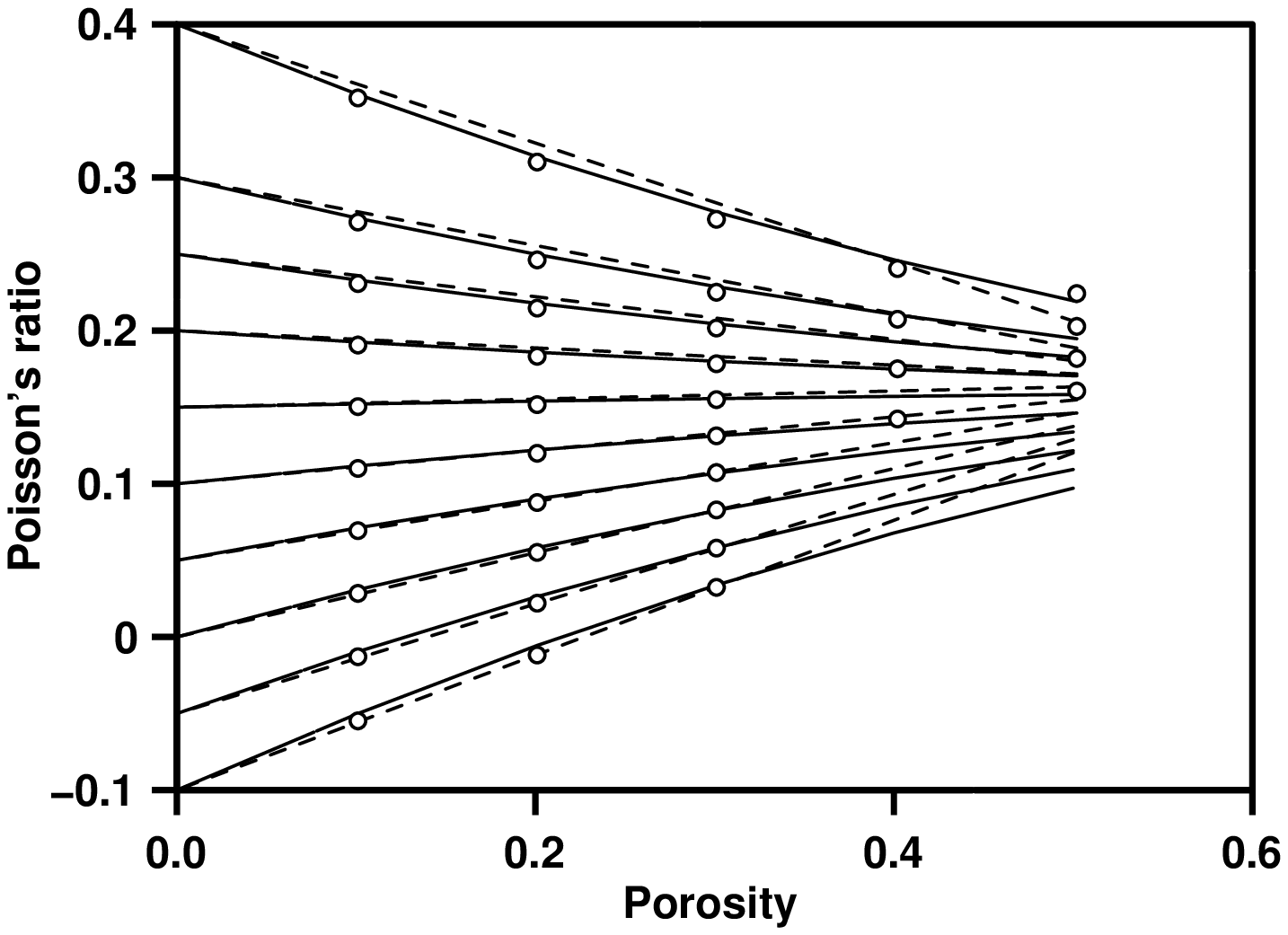}
\caption{Poisson's ratio of the overlapping ellipsoidal pore model
as a function of solid Poisson's ratio and porosity.
The solid lines are an empirical fit to Eq.~(\ref{percSform}) and the dashed
lines correspond to the linear fit to Eq.~(\ref{lineSform}) with $\nu_0=0.166$
and $\phi_0=0.604$.  The intercepts of the lines at zero porosity
correspond to the solid Poisson's ratio. 
\label{e1125_SvP}
}
\end{figure}

\noindent
{\bf Overlapping Spherical Pores.}
The overlapping spherical pore (or swiss cheese) 
model~\cite{TorqRev91,Weissberg63}
is generated
by interchanging the roles of the solid and pore phase of the
overlapping solid sphere model (see Fig.~\ref{gath3D}b).
The morphology corresponds to
isolated spherical pores at low porosity, with the pores becoming
macroscopically interconnected at $\phi\approx0.3$.
The solid phase remains connected up to $\phi\approx0.97$. This type
of morphology may arise in ceramics generated with a particulate
filler~\cite{Coble_K56}, or where bubbles form in a molten
state~\cite{Walsh_Glass65}.
We consider solid Poisson's ratios in the
range $-0.1 \leq \nu_s \leq 0.4$. 

We determined that
statistical errors were acceptable for a computational
cube of size $T$=12 $\mu$m with pores of radii $r$=1 $\mu$m. Using
$M$=80 pixels, the discretization errors were less than 3 \% for $\phi=0.5$
and 2 \% for $\phi=0.3$. 
Therefore, it was not considered necessary to generate
samples at different discretizations ($M$) and extrapolate the results.
As for solid spheres, the Young's modulus was independent
of the solid Poisson's ratio to a very good approximation. 
The Young's modulus can be described
by Eq.~(\ref{formE}) with $n=1.65$ and $\phi_0$=0.818
(Fig.~\ref{all_Evp}).  Poisson's ratio of the porous material is shown in
Fig.~\ref{E111_SvP} and is simply described by the linear relation 
given in Eq.~(\ref{lineSform}) with $\nu_0$=0.221
and $\phi_0$=0.840. Again, a flow diagram is observed.

\noindent
{\bf Overlapping Ellipsoidal Pores.}
A common method of analyzing the effect of pore shape
on elastic properties is to study ellipsoidal pores. In analytic
formulae, it is possible to treat the limiting cases of needles and platelets,
although the difficulty of resolving these fine structures
prohibits these limits from being treated with the
finite element method. However, the percolation properties of these
limiting cases can be computationally studied~\cite{GarbocziEllip95}.
To gauge the effect of deviations from spherical shaped pores
we considered isotropically oriented overlapping oblate ellipsoidal pores
bounded by the surface $(x/a)^2+(y/b)^2+(z/c)^2=1$
with $a$=$b$=1 $\mu$m and $c$=0.25 $\mu$m (see Fig.~\ref{gath3D}c).
For this case,
the pore phase becomes connected at porosity
$\phi=0.2$~\cite{GarbocziEllip95}.
Statistical errors were found to be acceptable for a computational
cube of size $T$=10 $\mu$m. Using $M$=96 pixels the discretization
errors were 3 \% for $\phi=0.5$ and 2 \% for $\phi=0.3$. 
As for the case
of spherical pores, these errors were considered sufficiently small, so that
the added computational burden of the extrapolation technique could be
again avoided.

Again the Young's modulus was found to be independent of
the solid Poisson's ratio to a very good approximation. 
The results,
shown in Fig.~\ref{all_Evp}, can be accurately described by
Eq.~(\ref{formE}) with $n$=2.25 and $\phi_0$=0.798.
The Poisson's ratio (Fig.~\ref{e1125_SvP}) can be roughly fit
using the Eq.~(\ref{lineSform}) with $\nu_0$=0.166 and $\phi_0$=0.604.
A better fit is obtained using Eq.~(\ref{percSform}) with
$m$=1.91, $\nu_0$=0.161 and $\phi_0$=0.959. A flow diagram similar to
that seen before is obtained.

The CPU time and memory required for these computations are an important
``experimental" detail.
The memory requirement for a given model was
$230 \times M^3$ bytes, where $M$ was the edge length in pixels of a cubic
unit cell. So for the largest computations carried out, $M = 128$, the memory
requirement was about $500$ Mbytes. The amount of CPU time consumed was 
approximately $3000$ hours, divided among different modern workstations.

\begin{center}\noindent{\bf III. Comparison with micro-mechanical and MSA formulae}\end{center}

\noindent
In this section we compare a selection of
well known theoretical results with the `numerically exact' data computed
in the previous section. These results include both analytically exact
results (bounds, expansions, dilute limits, composite sphere assemblage),
as well as approximate results (effective medium theories, minimum solid
area models). 

There are several kinds of exact bounds that have been derived for elastic
materials~\cite{Hashin83,TorqRev91}.
These are equations involving the different phase moduli, the 
volume fractions of the different phases, and various correlation functions
that define the geometry of the composite.  The upper bound gives the 
maximum possible composite elastic moduli, and the lower bound gives the 
lowest possible composite elastic moduli. 
The bounds used in this paper are three-
point bounds, which have been written out explicitly for overlapping solid spheres
and overlapping spherical pores~\cite{TorqRev91}. In the case where
one phase has zero elastic moduli, as is true in this paper, the lower bound
becomes zero as well, and so only the upper bound is meaningful.

An exact perturbation expansion also exists, where the elastic moduli 
of a two-phase 
material is expanded in terms of parameters involving the individual 
elastic moduli of each phase and geometrical
quantities~\cite{Torqpower,TorqpowerII}. 
This expansion has been carried out to three terms explicitly, and it is 
this truncated form to which we will compare our numerical data.
The result is expected to be accurate when the void phase is not
interconnected.

Another exact result, which is used later in this section to build the 
various effective medium theories, is the case of
dilute spherical pores for which the exact effective moduli are given by
\begin{eqnarray}
\label{theory_dilK}
K&=&K_m + c_i P^{mi} (K_i-K_m) \\
G&=&G_m + c_i Q^{mi} (G_i-G_m)
\label{theory_dilG}
\end{eqnarray}
with
\begin{eqnarray}
P^{mi}&=&\frac{3K_m+4G_m}{3K_i+4G_m},\;\; \nonumber
Q^{mi}=\frac{G_m+F_m}{G_i+F_m}, \\
F_m&=&\frac{G_m}6 \frac{9K_m+8G_m}{K_m+2G_m}.
\label{theory_dilPQ}
\end{eqnarray}
Here $c_i$ denotes the concentration (volume fraction) of
inclusions and the subscripts $i$ and $m$ on the bulk $K$ and shear
modulus $G$ denote the properties of the inclusion and matrix,
respectively.
The result is attributed to numerous authors~\cite{Hashin83}.
For a porous matrix $K_i=G_i=0$ and the porosity is 
$\phi=c_i$.
The result is strictly valid for small concentrations of inclusions
$\phi\ll 1$ (in practice $\phi < 0.1$.) Cast in terms of the
engineering constants for porous inclusions this result becomes
\begin{eqnarray}
\label{ERoberts}
E&=&E_m - \frac32 \phi E_m \frac{9-4\nu_m-5\nu_m^2}{7-5\nu_m} + O(\phi^2) \\
\nu&=&\nu_m - \frac32 \phi \frac{(5\nu_m-1)(1-\nu_m^2)}{7-5\nu_m} + O(\phi^2) 
\end{eqnarray}

Our prior statement\cite{Roberts_TAg99} of Eq.~(\ref{ERoberts}) 
inadvertently omitted the factor of $3/2$,
although the correct result was used in the paper.
A non zero quadratic term can be added (as an empirical correction)
to ensure that $E$=0 at $\phi$=1.
This was suggested by Coble and Kingery~\cite{Coble_K56}
for MacKenzie's~\cite{MacKenzieHoles}
result for spherical pores, which is equivalent to
Eqs.~(\ref{theory_dilK})-(\ref{theory_dilPQ})
with $K_i=G_i=0$. 

To adapt the dilute formulas to the case of finite 
porosity a number
of proposals have been made. The approximate equations that result are usually
called effective medium theories. The most common approximation
is the so-called self consistent method (SCM) of Hill~\cite{HillSCM}
and Budiansky~\cite{BudianskySCM}. In
this model the equations of elasticity are solved
for a spherical inclusion embedded
in a medium of unknown effective moduli. The effective
moduli $K$ and $G$ are then derived. In the dilute case the embedding
medium is just the matrix.  The Hill-Budiansky result can be stated
as~\cite{Berryman80_ell} 
\begin{eqnarray}
c_i P^{*i}(K_i-K_*) + c_m P^{*m}(K_m-K_*)&=&0 \\
c_i Q^{*i}(G_i-G_*) + c_m Q^{*m}(G_m-G_*)&=&0 
\end{eqnarray}
where $K_*$ and $G_*$ denote the effective moduli and $P^{*m}$ and $Q^{*m}$
are given in Eq.~(\ref{theory_dilPQ}). The equations cannot
be explicitly solved and numerical methods are necessary
(see Hill~\cite{HillSCM} and Berryman~\cite{Berryman80_ell} for details).
In the case of porous inclusions,
the moduli vanish at $\phi=\frac12$, which is a 
property not shared with most composites
(e.g.,\ the overlapping sphere model). To derive a more realistic
result, Christensen and Lo~\cite{ChristensenLoGSCM} generalized the SCM (GSCM)
to the case of a spherical shell embedded in a matrix of
unknown moduli. The result is complicated and not reproduced here.

The differential method (reviewed by McLaughlin~\cite{McLaughlinDM77})
provides an alternative model using a similar philosophy. Suppose that the
effective moduli of a composite medium are known to be $K_*$ and $G_*$.
Now if a small additional concentration of inclusions are added, the change
in $K_*$ and $G_*$ is approximated to be that which would arise if a
dilute concentration of inclusions were added to a uniform, homogeneous 
matrix with
moduli $K_*$ and $G_*$. This leads to a pair of coupled differential
equations,
\begin{eqnarray}
\label{DMK}
\frac{dK_*}{d c_i}&=&P^{*i} \frac {K_i-K_*}{1-c_i};\; K_*(c_i=0)=K_m \\
\frac{dG_*}{d c_i}&=&Q^{*i} \frac {G_i-G_*}{1-c_i};\; G_*(c_i=0)=G_m.
\label{DMG}
\end{eqnarray}

The dilute result, the self consistent
result~\cite{Berryman80_ell}, and the differential method~\cite{McLaughlinDM77}
can all be extended to the case of spheroidal
inclusions. The general results~\cite{Wu66} for $P^{mi}$ and $Q^{mi}$
have been given by
Berryman~\cite{Berryman80_ell}. In addition
to these results Wu~\cite{Wu66} derived
a variant of the self consistent method, where $K_*$ and $G_*$, the effective
moduli, 
are found by implicitly solving
the equations
\begin{eqnarray}
\label{theory_WuK}
K_*&=&K_m + c_i P^{*i} (K_i-K_m) \\
G_*&=&G_m + c_i Q^{*i} (G_i-G_m).
\label{theory_WuG}
\end{eqnarray}

A different type of microstructure is provided by
Hashin's~\cite{Hashin62CSA,Hashin83} model of space-filling poly-disperse
hollow spheres (the ``composite-sphere assemblage''). Although
a simple formula exists for the bulk modulus over the full porosity
range~\cite{Hashin62CSA}, exact results for the Young's moduli
are not available.  Ramakrishnan and Arunachalam~\cite{RamAru90} recently
derived the approximation
\begin{eqnarray}
\frac{E}{E_s}&=&\frac{(1-\phi)^2}{(1+2\phi-3\nu_s\phi)} \label{RAE} \\
\nu&=& \frac{(4\nu_s+3\phi-7\nu_s\phi)}{4(1+2\phi-3\nu_s\phi)}. \label{RAS} 
\end{eqnarray}
However, the derivation is not rigorous.  In particular, the
exact result for the bulk modulus of the model~\cite{Hashin62CSA} is
around twice that predicted by Eqs.~(\ref{RAE}--\ref{RAS}) at $\phi=0.5$.
Since Eq.~(\ref{RAE}) was found to provide
reasonable agreement with experimental data for porous
ceramics~\cite{RamAru93}, we compare its predictions to our FEM data below.

\begin{figure}[t]
\centering \epsfxsize=8.5cm \epsfbox{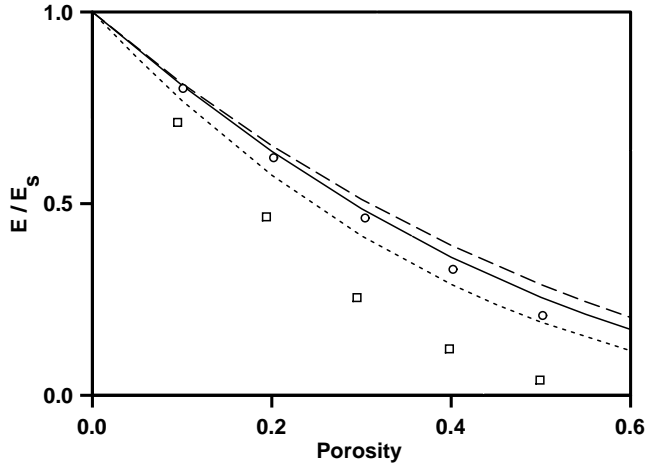}
\caption{
A comparison of rigorous bounds and expansions to the FEM data for 
overlapping spherical pores
($\circ$) and overlapping solid spheres ($\Box$).  
The truncated expansion (------) and the three-point upper bound (-- -- --) 
are shown for the spherical pore case.  Only the three-point bound
($\cdots$) is shown for the solid sphere case. The three-point lower bound 
is zero for porous materials. The Poisson's ratio is 0.2 for all the results.
\label{e111vBnds}
}
\end{figure}

The final class of results we consider is provided by the `minimum solid
area' (MSA) models~\cite{Knudsen_MSA}
(which have been recently reviewed by
Rice~\cite{Rice_Eval_MSA,Rice_Comp_MSA}).
This approach is based on the assumption
that the ratio of the effective moduli to the solid moduli
is directly proportional to the minimum ratio of
solid contact area to the total cross-sectional area of
periodic structures. The approximation derived depends
on the particular model considered. We consider two basic
models most closely aligned with our FEM data: simple cubic
arrays of solid and porous spheres. The latter
case provides a particularly simple example of the
type of result which can be derived. Suppose the
repeat distance of the lattice is $2h$ and the sphere radius
is $r$. The Young's modulus is assumed to be proportional
to the area fraction, giving
\begin{eqnarray}
\frac{E}{E_s}=\frac{(2h)^2-\pi r^2}{(2h)^2}=1-\frac\pi4
\left( \frac{6}{\pi} \right)^{\frac23} \phi^\frac23
\end{eqnarray}
since $\phi=\frac16 \pi (r/h)^3$. The form of the result
changes for $r>h$ (or $\phi>\pi/6$=0.52) as the spheres begin
to coalesce. Rice~\cite{Rice_Eval_MSA} has noted the results of many
different periodic structures can be approximated by the
form $E/E_s=e^{-b\phi}$ over a range of porosities. For example,
$b\approx5$ for the solid sphere model and $b\approx3$ for the
porous sphere model. It is argued that for a given
set of data, $b$ can be compared with known values
to assess the type of porosity. Often fractions
of different types of porosity are assumed to match
experimental data making the method an interpretive rather
than a predictive tool.  Since we have measured $E$ for
microstructures based on solid sphere contacts and porous
spheres we should be able to ascertain the accuracy of the
MSA formulae for these cases.

Fig.~\ref{e111vBnds} shows the comparison between the exact three-point 
bounds~\cite{TorqRev91} for the overlapping solid sphere and
spherical pore case, the 
truncated expansion~\cite{Torqpower,TorqpowerII}
for the overlapping spherical pore case, and the 
numerical results. Clearly the expansion does better than the three-point
bound for the overlapping spherical pore case, though both are fairly close
to the numerical results.  The bound lies far away from the overlapping
solid sphere numerical results, however.  For this case, the truncated
expansion does not exist. Only the $\nu_s = 0.2$ data is shown. Using 
the truncated expansion, one can show that in 3-D, the Young's modulus is not
exactly independent of the solid Poisson's ratio, but is rather a very good
approximation, as was shown earlier in this paper.

\begin{figure}[t]
\centering \epsfxsize=8.5cm \epsfbox{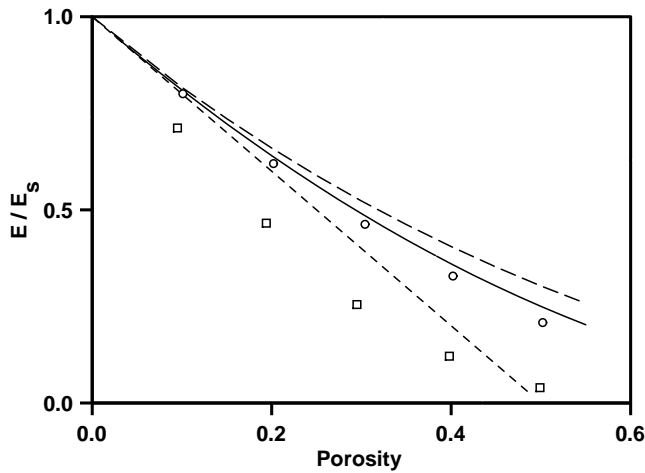}
\caption{
A comparison of different theories to the FEM data for overlapping spherical
pores ($\circ$).  The lines correspond to the dilute result and
self consistent method~\protect\cite{HillSCM,BudianskySCM} (or SCM) (- - -), the
differential method~\protect\cite{McLaughlinDM77} and dilute result with Coble--Kingery correction
(------) and the generalized SCM~\protect\cite{ChristensenLoGSCM} (--- --- ---).
Data for the overlapping solid sphere model (for which no rigorous
theories exist) are also shown ($\Box$).
\label{e111_EvT}
}
\end{figure}

In Fig.~\ref{e111_EvT}, we compare the FEM data ($\nu_s=0.2$)
for overlapping spherical pores with dilute and effective medium theory 
analytic
results. At this Poisson's ratio the SCM and dilute results reduce
to $E/E_s=1-2\phi$ while the differential and dilute results with the
Coble--Kingery correction reduce to $E/E_s=(1-\phi)^2$.
Since the analytic results are
based on the case of dilute spherical pores
they all match the FEM data at $\phi=0.1$.
The dilute and SCM results under-estimate the FEM data at higher
porosities because of the aphysical percolation threshold at
$\phi=\frac12$. The generalized SCM over-estimates the data,
while the differential method performs reasonably well over
the entire porosity range. The latter observation might have been
anticipated given the close association between
the definition of the spherical pore model and the assumption
of the differential method. At increasing porosities
we are simply adding additional spherical pores to a
porous matrix. The data for overlapping solid spheres
is also shown in the figure, and seen to be quite different
from any of the available results. This demonstrates that microstructure
(the geometrical nature of the porosity) is an important factor
besides the actual value of the porosity.

\begin{figure}[t]
\centering \epsfxsize=8.3cm\epsfbox{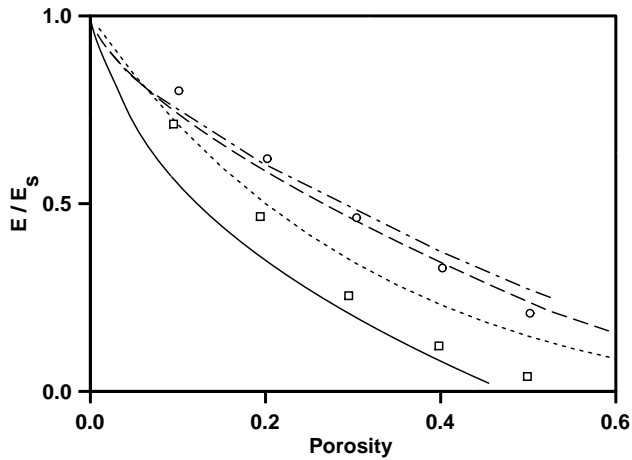}
\caption{
A comparison of the minimum solid area (MSA)
models~\protect\cite{Rice_Eval_MSA}
to the FEM data for overlapping spherical pores ($\circ$) and
solid spheres ($\Box$).
The MSA solid sphere model (------) and MSA porous sphere model
(-- -- --) (in simple cubic packings) are seen to under-estimate the
data for low porosities ($\phi<0.3$). 
The formula of
Ramakrishnan and Arunachalam~\protect\cite{RamAru90}
$E/E_s=(1-\phi)^2/(1+1.4\phi)$ ($\cdots$)
and the results of the
generalized method of cells for a periodic spherical
pore~\protect\cite{HerBax99} (-- $\cdot$ --) are also shown.
\label{EvsMSA}
}
\end{figure}

\begin{figure}[ht!]
\centering \epsfxsize=8.3cm\epsfbox{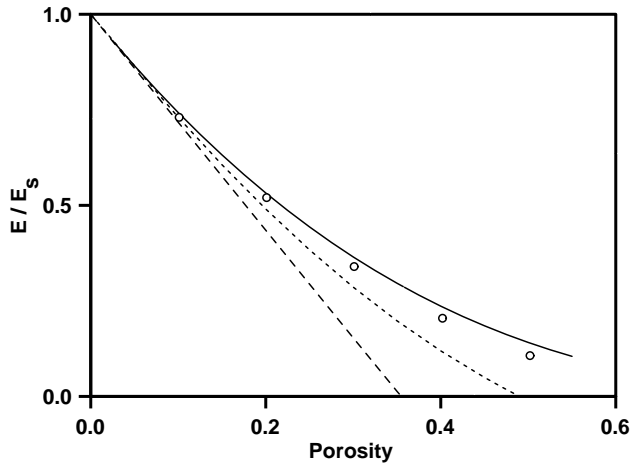}
\caption{
A comparison of different theories to the FEM data for overlapping oblate
ellipsoidal pores ($\circ$).  The lines correspond to the differential
method~\protect\cite{McLaughlinDM77} (------)
and the self consistent methods of Wu~\protect\cite{Wu66}
(-- -- --) and Berryman~\protect\cite{Berryman80_ell} ($\cdots$).
\label{e1125_EvT}
}
\end{figure}

In Fig.~\ref{EvsMSA}, the minimum solid area models and the
Ramakrishnan and Arunachalam results~\cite{RamAru90} are compared with the data.
The MSA model for spherical pores performs reasonably well, although
underestimating the FEM data for overlapping random spherical pores
at low porosities $\phi<0.3$. The MSA model for solid spheres
considerably underestimates these data for $\phi<0.3$. The
Ramakrishnan and Arunachalam~\cite{RamAru90} approximation falls
between the FEM data for $\phi>0.1$ indicating that it corresponds to neither
of the microstructures. For purposes of comparison we also
report numerical results obtained using the computational
generalized method of cells~\cite{HerBax99}.
For a periodic spherical pore the results significantly
underestimate the FEM data for overlapping spherical pores at low
porosities (and hence the exact dilute result).
It is not clear if this is due to the assumptions,
or the particular implementation, of the method.

The FEM data for overlapping
oblate ellipsoidal pores is compared with the available theories
in Fig.~\ref{e1125_EvT}. The SCM results of Wu~\cite{Wu66} and
Berryman~\cite{Berryman80_ell} underestimate the porosity as
a result of underestimating the physical percolation threshold.
The Berryman result performs significantly better than does the Wu result.
As for the case of spheres, the
differential method matches the data quite closely because of
the similarity between the assumptions of the theory and the
definition of the model.

We have also compared the Poisson's ratio predicted by
the various self-consistent and differential methods to the FEM
data for overlapping spherical and ellipsoidal pores. The theoretical
results converge to different fixed points (e.g.\ Fig.~\ref{oss_SvP})
in qualitative agreement with the data. But only the differential method
provides reasonable agreement with the FEM data (with absolute error less than
0.02 for $\phi\leq0.4$ and $0.1 \leq \nu_s \leq 0.4$).

\begin{center}\noindent{\bf IV. Comparison with Experiment}\end{center}

\begin{figure}[t]
\centering \epsfxsize=8.3cm\epsfbox{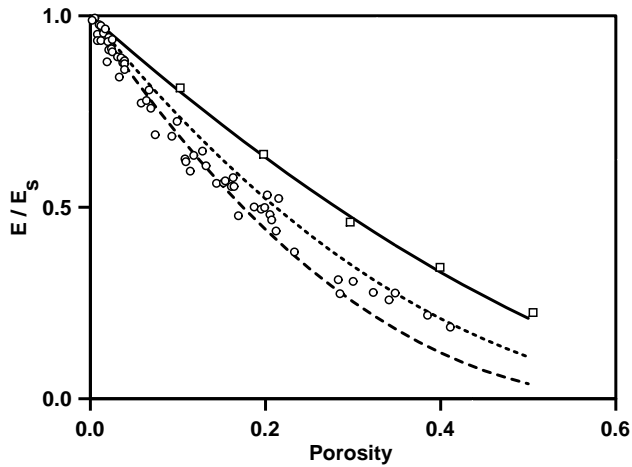}
\caption{
Data for alumina ($E_s$=410 GPa) compiled by Knudsen~\protect\cite{Knudsen_Al62}
($\circ$).
The Coble-Kingery~\protect\cite{Coble_K56} ($E_s$=386 GPa)
data are also shown ($\Box$).
The lines correspond to the FEM theories computed in this paper:
overlapping spherical pores (------),
overlapping oblate ellipsoidal pores ($\cdots$) and
overlapping solid spheres (-- -- --).
\label{alumina}
}
\end{figure}

\noindent
We now use the FEM results to analyze experimental measurements of the elastic
properties of porous ceramic materials.
The dependence of the elastic moduli on porosity has been the subject
of many studies
~\cite{Rice_Comp_MSA,RamAru93,Bocc94}.
Data for porous
alumina from numerous studies
~\cite{Knudsen_Al62}
are shown in Fig.~\ref{alumina}. The Coble-Kingery~\cite{Coble_K56}
material is markedly
stiffer than other materials, and is in very good agreement with
the FEM results for the overlapping spherical pore model. The pores in
the alumina matrix were actually created by the incorporation of
a particulate filler~\cite{Coble_K56}, which corresponds well with
the definition of the model microstructure. 
The remaining data closely follow
the overlapping solid sphere FEM result for $\phi<0.25$,
indicating that the solid alumina phase has the sintered granular
morphology exhibited by the model microstructure (Fig.\ref{gath3D}a).
However, Knudsen notes that several of the samples summarized
were also created using particulate fillers. At higher porosities
the solid sphere result underestimates the data. One reason for this
might be that the model contains isolated solid spheres which
artificially reduce the actual porosity. This was checked and found not
to be the case for the porosities studied. Therefore, the solid
connections in these samples of porous alumina are likely stiffer
than those found in the solid sphere model at porosities $\phi>0.25$.
Overlapping spheres can create very sharp ``valleys" between a pair
of overlapping solid spheres (see Fig.~\ref{gath3D}a),
which would be 
rounded off in the sintering process, presumably strengthening the solid-solid
connection.

\begin{figure}[t]
\centering \epsfxsize=8.3cm\epsfbox{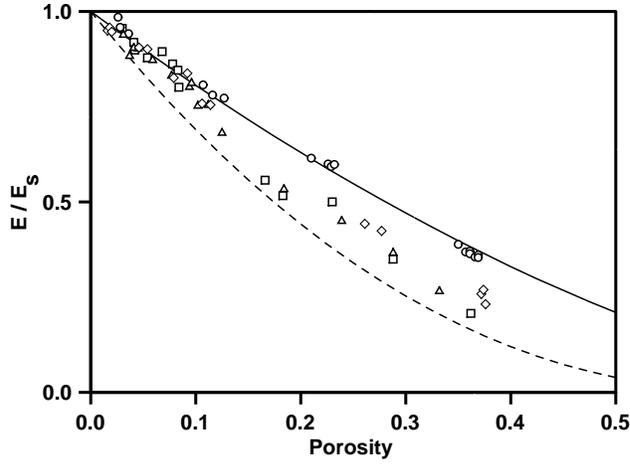}
\caption{
Data for various oxides measured by Hunter {\em et al}
~\cite{Hunter_sm2o3,Hunter_lu2o3,Hunter_gd2o3,Hunter_hfo2}
compared with the FEM theories for overlapping spherical pores (------),
and overlapping solid spheres (-- -- --).
Sm$_2$O$_3$~\protect\cite{Hunter_sm2o3} $E_s$=145 GPa ($\diamond$);
Lu$_2$O$_3$~\protect\cite{Hunter_lu2o3} $E_s$=193 GPa ($\Delta$);
Gd$_2$O$_3$~\protect\cite{Hunter_gd2o3} $E_s$=150 GPa ($\circ$);
HfO$_2$~\protect\cite{Hunter_hfo2} $E_s$=246 GPa ($\Box$);
\label{hunter}
}
\end{figure}

Hunter 
{\em et al.}~\cite{Hunter_sm2o3,Hunter_lu2o3,Hunter_gd2o3,Hunter_hfo2}
have studied the Young's modulus of several different oxides. In all
cases, the porous material was created by sintering a powder of
the pure oxide. The results for the Young's modulus are reproduced
in Fig.~\ref{hunter}.  For low porosities ($\phi<0.1$) all
of the data followed the FEM results for overlapping spherical
pores.
For Gd$_2$O$_3$ the FEM result continues to provide excellent
agreement up to the maximum porosity measured ($\phi=0.4$)
indicating that the microstructure is similar to that of the model
(overlapping pores).  In contrast, the data for the other three
oxides decreases towards the result for overlapping solid spheres
indicating a more granular character. 

\begin{figure}[hbt]
\centering \epsfxsize=8.3cm\epsfbox{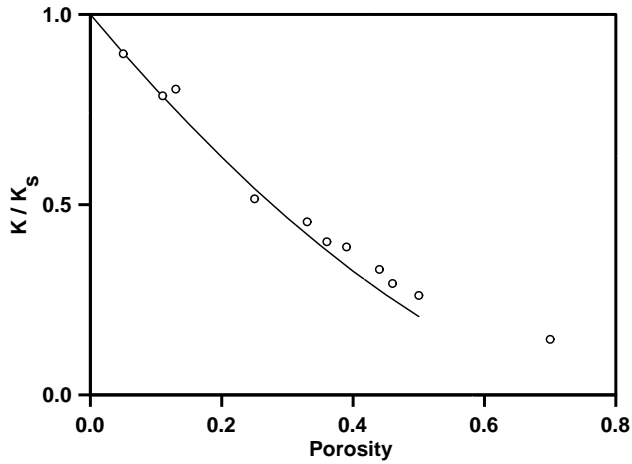}
\caption{
Data for porous glass~\protect\cite{Walsh_Glass65}
($K_s$=46 GPa, $\nu_s$=0.23).
The line corresponds to the FEM theory for
overlapping spherical pores (------).
\label{walsh65}
}
\end{figure}

\begin{figure}[t]
\centering \epsfxsize=8.3cm\epsfbox{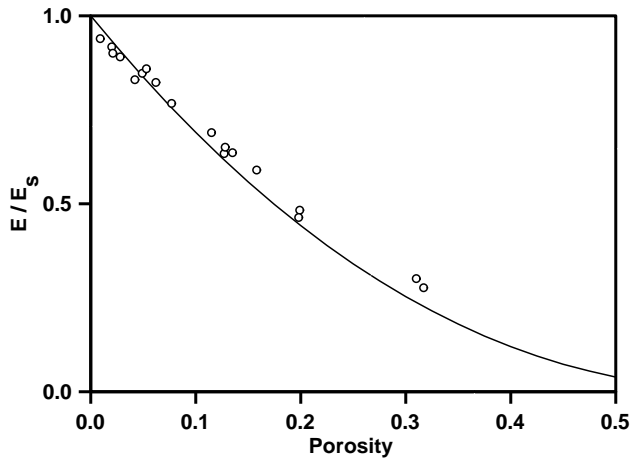}
\caption{
Data for MgAl$_2$O$_4$~\protect\cite{Porter_spinel}.
We used the value
$E_s$=41.2$\times10^{6}$psi (284 GPa) 
indicated on Fig.~3(A) of
the reference, rather than the reported value of
$E_s$=43.4$\times10^{6}$psi which appears to be a misprint.
The line corresponds to the FEM theory for 
overlapping solid spheres.
\label{porter}
}
\end{figure}

The data of Walsh {\em et al.}~\cite{Walsh_Glass65} for
porous glass is compared
with the FEM results for overlapping spherical pores
in Fig.~\ref{walsh65}. The agreement is good for small
to moderate porosities ($\phi < 0.3$), but the FEM results
underestimate the data at higher porosities. 
Walsh {\em et al.} point out that the pores in the glass are
actually not interconnected (unlike the overlapping pores of the model).
This would account for the increased stiffness.
It is interesting that the FEM results begin to deviate from
the experimental data at the threshold where the pores become
macroscopically connected ($\phi$=0.3). 
Data for sintered MgAl$_2$O$_2$~\cite{Porter_spinel} powder
is shown in Fig.~\ref{porter}, and is well modeled by the
FEM results for overlapping solid spheres.
Micrographs of the ceramic indicate a granular structure similar to
that of the model microstructure (although the grains appear more
like polyhedra, not spheres). 

\clearpage
 
\begin{center}\noindent{\bf V. Discussion and Conclusions}\end{center}

\noindent
We have derived empirical theories for the dependence of the
Young's modulus on porosity for three
distinct models of porous ceramics, based on careful finite element
computations. 
An advantage of these results over many conventional theories is
that they correspond to {\em a priori} known microstructures.
The dilute result (extended by Coble-Kingery to all porosities),
the differential method, and the self consistent methods 
all have a ``built-in'' microstructure, but apart from the
dilute case, it is not clear what that microstructure is.
Therefore, agreement or disagreement with a particular
analytic theory neither confirms nor rejects a particular
physical model. For the minimum solid area models, the microstructure
is exactly known, but the approximation involved in making the
Young's modulus directly proportional to the contact area leads
to a similar conclusion. Indeed, we found that the MSA models did
not provide quantitative agreement with the moduli of
the random microstructures studied. 
We found that the differential method
(Eqs.~\ref{DMK} and~\ref{DMG}) gave results
in reasonable agreement with computed data for the
cases of overlapping spherical and ellipsoidal pores,
probably due to the similarities
between the assumptions of the model and the definitions
of the microstructure. Results for the granular model of
overlapping solid spheres were not well modeled
by any of the analytic theories, demonstrating the importance
of finite element techniques in this case of great physical interest.

We have also generated data that shows the dependence of Poisson's
ratio on porosity and the solid Poisson's ratio. It is difficult
to study this question experimentally because of the inability to
vary the Poisson's ratio of the solid independently, and the well
known difficulties of accurately measuring the Poisson's ratio
at moderate to high porosities~\cite{Bocc94}. At sufficiently
high porosities 
we find that the Poisson's ratio converges to a fixed non-zero
value $\nu_0$ irrespective of the
solid Poisson's ratio. For overlapping solid spheres $\nu_0=0.14$,
spherical pores $\nu_0=0.22$ and oblate ellipsoidal pores $\nu_0=0.16$.
This behavior is exact in
two dimensions~\cite{Garboczi_Circ_Holes92,Cherk92} 
and is exhibited by many of the analytic theories in three dimensions.
At present the available experimental data cannot confirm this
qualitative behavior~\cite{Bocc94}. We have shown that the
Poisson's ratio does
not vanish at high porosities as has been recently
argued~\cite{Rice_Comm_PR}.

It is not simple to attribute our results to features of the solid-pore
morphology such as the size, shape, distribution and connectivity
of pores or solid grains, since these features
have no obvious definition for complex bi-continuous random
microstructures.  A few general observations can be made, and interpreted
in terms of interrelated geometrical and mechanical features of the
models.
For a given porosity, the sintered grain structure
of the overlapping solid sphere model is relatively weak. The small
solid contacts between spheres and the highly interconnected
porosity (which becomes macroscopically connected at $\phi$=0.03)
lead to a weak structure. We also assume that the
valleys which occur between grains will provide sites of
large stress concentrations, and consequently, large deformations.
In contrast, spherical pores provide high (near optimal) stiffness at a
given
porosity.  The dispersed nature of the porosity (which is macroscopically
disconnected for $\phi <0.3$) corresponds to a well connected
solid matrix. Ellipsoidal pores tend to weaken a structure more than
spherical pores due to a combination of a less well connected solid phase
(the pores become macroscopically connected at $\phi=0.2$),
and greater stresses and deformations near the high curvature
regions of the ellipsoid.

We have compared our FEM results with several sets of previously
published experimental data. In cases where the microstructure of the porous
ceramics roughly matched that of the models, the agreement
was very good. Since the FEM results correspond to a known
microstructure, it was possible to explain deviations
in terms of specific microstructural features. Thus, comparison
of experimental data with the three computational results 
provides a useful interpretive
tool. Note that a given elastic modulus does not correspond to
a particular microstructure.
Therefore, it is important to corroborate microstructural interpretations
obtained from the elastic moduli with information about
the particular material (such as a micrograph).
In the future it would be useful to extend this work
to higher porosities and to other relevant models
(such as non-overlapping porous spheres).
It is also possible to use statistical microstructural information
obtained from two-dimensional micrographs to generate
models~\cite{Roberts_TAg99} that actually mimic physical
microstructures. 

\noindent
{\bf Acknowledgments}
A.R.\ thanks the Fulbright Foundation and Australian Research Council
for financial support. We also thank the Partnership for
High-Performance Concrete Technology program of the National Institute
of Standards and Technology for partial support of this work.


\begin{thebibliography}{10}
\newcommand{\enquote}[1]{``#1''}

\bibitem{Coble_K56}
R.~L. Coble and W.~D. Kingery, \enquote{Effect of Porosity on Physical
  Properties of Alumina}, {\em J. Am. Ceram. Soc.\/}, {\bf 39}(11), 377--385
  (1956).

\bibitem{DeanLopez83}
E.~A. Dean and J.~A. Lopez, \enquote{Empirical Dependence of Elastic Moduli on
  Porosity for Ceramic Materials}, {\em J. Am. Ceram. Soc.\/}, {\bf 60}(7-8),
  345--349 (1977).

\bibitem{Rice_Eval_MSA}
R.~W. Rice, \enquote{Evaluation and extension of physical property-porosity
  models based on minimum solid area}, {\em J. Mater. Sci.\/}, {\bf 31},
  102--118 (1996).

\bibitem{HerBax99}
C.~T. Herakovich and S.~C. Baxter, \enquote{Influence of pore geometry on the
  effective response of porous media}, {\em J. Mater. Sci.\/}, {\bf 31},
  1595--1609 (1999).

\bibitem{Hashin83}
Z.~Hashin, \enquote{Analysis of composite-materials - a survey}, {\em J. Appl.
  Mech.\/}, {\bf 50}, 481--505 (1983).

\bibitem{AboudiBook}
J.~Aboudi, Mechanics of composite materials: a unified micromechanical
  approach, Elsevier, Amsterdam, 1991.

\bibitem{ChristensenBook}
R.~M. Christensen, Mechanics of composite materials, Wiley, New York, 1979.

\bibitem{McLaughlinDM77}
R.~McLaughlin, \enquote{A study of the differential scheme for composite
  materials}, {\em Int. J. Eng. Sci.\/}, {\bf 15}, 237--244 (1977).

\bibitem{HillSCM}
R.~Hill, \enquote{A self-consistent mechanics of composite materials}, {\em J.
  Mech. Phys. Solids\/}, {\bf 13}, 213--222 (1965).

\bibitem{BudianskySCM}
B.~Budiansky, \enquote{On the elastic moduli of some heterogeneous materials},
  {\em J. Mech. Phys. Solids\/}, {\bf 13}, 223--227 (1965).

\bibitem{Wu66}
T.~T. Wu, \enquote{The effect of inclusion shape on the elastic moduli of a
  two-phase material}, {\em Int. J. Solids Structures\/}, {\bf 2}, 1--8 (1966).

\bibitem{Berryman80_ell}
J.~G. Berryman, \enquote{Long-wavelength propagation in composite elastic media
  II. Ellipsoidal inclusions}, {\em J. Acoust. Soc. Am.\/}, {\bf 68}(6),
  1820--1831 (1980).

\bibitem{Rice_Comp_MSA}
R.~W. Rice, \enquote{Comparison of physical property-porosity behaviour with
  minimum solid area models}, {\em J. Mater. Sci.\/}, {\bf 31}, 1509--1528
  (1996).

\bibitem{TorqRev91}
S.~Torquato, \enquote{Random heterogeneous media: Microstructure and improved
  bounds on effective properties}, {\em Appl. Mech. Rev.\/}, {\bf 44}, 37--76
  (1991).

\bibitem{Garboczi95a}
E.~J. Garboczi and A.~R. Day, \enquote{An algorithm for computing the effective
  linear elastic properties of heterogeneous materials: Three-dimensional
  results for composites with equal phase Poisson ratios}, {\em J. Mech. Phys.
  Solids\/}, {\bf 43}, 1349--1362 (1995).

\bibitem{Adlerelas1}
J.~Poutet, D.~Manzoni, F.~Hage-chehade, C.~G. Jacquin, M.~J. Bouteca, J.~F.
  Thovert, and P.~M. Adler, \enquote{The effective mechanical properties of
  random porous media}, {\em J. Mech. Phys. Solids\/}, {\bf 44}, 1587--1620
  (1996).

\bibitem{RamAru93}
N.~Ramakrishnan and V.~S. Arunachalam, \enquote{Effective Elastic Moduli of
  Ceramic Materials}, {\em J. Am. Ceram. Soc.\/}, {\bf 76}(11), 2745--52
  (1993).

\bibitem{Bocc94}
A.~R. Boccaccini, \enquote{Comment on ``Effective Elastic Moduli of Ceramic
  Materials''}, {\em J. Am. Ceram. Soc.\/}, {\bf 76}(10), 2745--52 (1994).

\bibitem{Rice_Comm_PR}
R.~W. Rice, \enquote{Comment on ``Effective Elastic Moduli of Porous Ceramic
  Materials''}, {\em J. Am. Ceram. Soc.\/}, {\bf 78}(6), 1711 (1995).

\bibitem{Eds_manual}
E.~J. Garboczi (1998), ${\rm N}$IST Internal Report 6269, available at
  http://ciks.cbt.nist.gov/garboczi/, Chapter 2.

\bibitem{Weissberg63}
H.~L. Weissberg, \enquote{Effective diffusion coefficient in porous media},
  {\em J. Appl. Phys.\/}, {\bf 34}, 2636--2639 (1963).

\bibitem{Roberts95a}
A.~P. Roberts and M.~Teubner, \enquote{Transport properties of heterogeneous
  materials derived from Gaussian random fields: Bounds and simulation.}, {\em
  Phys. Rev. E\/}, {\bf 51}, 4141--4154 (1995).

\bibitem{Garboczi_Circ_Holes92}
A.~R. Day, K.~A. Snyder, E.~J. Garboczi, and M.~F. Thorpe, \enquote{The elastic
  moduli of sheet containing spherical holes}, {\em J. Mech. Phys. Solids\/},
  {\bf 40}, 1031--1051 (1992).

\bibitem{Cherk92}
A.~V. Cherkaev, K.~A. Lurie, and G.~W. Milton, \enquote{Invariant properties of
  the stress in plane elasticity and equivalence classes of composites}, {\em
  Proc. R. Soc. Lond. A\/}, {\bf 438}, 519--529 (1992).

\bibitem{Walsh_Glass65}
J.~B. Walsh, W.~F. Brace, and A.~W. England, \enquote{Effect of Porosity on
  Compressibility of Glass}, {\em J. Am. Ceram. Soc.\/}, {\bf 48}(12), 605--608
  (1965).

\bibitem{GarbocziEllip95}
E.~Garboczi, K.~Snyder, J.~Douglas, and M.~Thorpe, \enquote{Geometrical
  percolation threshold of overlapping ellipsoids}, {\em Phys. Rev. E\/}, {\bf
  52}, 819--828 (1995).

\bibitem{Torqpower}
S.~Torquato, \enquote{Effective stiffness tensor of composite media-I. Exact
  series expansions}, {\em J. Mech. Phys. Solids\/}, {\bf 45}, 1421--1448
  (1997).

\bibitem{TorqpowerII}
S.~Torquato, \enquote{Effective stiffness tensor of composite media-II.
  Applications to isotropic dispersions}, {\em J. Mech. Phys. Solids\/}, {\bf
  46}, 1411--1440 (1998).

\bibitem{Roberts_TAg99}
A.~P. Roberts and E.~J. Garboczi, \enquote{Elastic properties of a
  tungsten-silver composite by reconstruction and computation}, {\em J. Mech.
  Phys. Solids\/}, {\bf 47}(10), 2029--2055 (1999).

\bibitem{MacKenzieHoles}
J.~F. MacKenzie, \enquote{Elastic constants of a solid containing spherical
  holes}, {\em Proc. Phys. Soc. (London)\/}, {\bf 63B}(1), 2--11 (1960).

\bibitem{ChristensenLoGSCM}
R.~M. Christensen and K.~H. Lo, \enquote{Solutions for effective shear
  properties in three phase sphere and cylinder models}, {\em J. Mech. Phys.
  Solids\/}, {\bf 27}, 315--330 (1979).

\bibitem{Hashin62CSA}
Z.~Hashin, \enquote{The elastic moduli of heterogeneous materials}, {\em ASME
  J. Appl. Mech.\/}, {\bf 29}, 143--150 (1962).

\bibitem{RamAru90}
N.~Ramakrishnan and V.~S. Arunachalam, \enquote{Effective elastic moduli of
  porous solids}, {\em J. Mater. Sci.\/}, {\bf 25}, 3930--3937 (1990).

\bibitem{Knudsen_MSA}
F.~P. Knudsen, \enquote{Dependence of mechanical strength of brittle
  polycrystalline specimens on porosity and grain size}, {\em J. Am. Ceram.
  Soc.\/}, {\bf 42}(8), 376--387 (1959).

\bibitem{Knudsen_Al62}
F.~P. Knudsen, \enquote{Effect of Porosity on Young's Modulus of Alumina}, {\em
  J. Am. Ceram. Soc.\/}, {\bf 45}(2), 94--95 (1962).

\bibitem{Hunter_sm2o3}
O.~Hunter, H.~J. Korklan, and R.~R. Suchomel, \enquote{Elastic Properties of
  Polycrystalline Monoclinic Sm$_2$O$_3$}, {\em J. Am. Ceram. Soc.\/}, {\bf
  57}(6), 267--268 (1974).

\bibitem{Hunter_lu2o3}
O.~Hunter and G.~E. Graddy, \enquote{Porosity Dependence of Elastic Properties
  of Polycrystalline Lu$_2$O$_3$}, {\em J. Am. Ceram. Soc.\/}, {\bf 59}(1-2),
  82 (1976).

\bibitem{Hunter_gd2o3}
J.~A. Haglund and O.~Hunter, \enquote{Elastic Properties of Polycrystalline
  Monoclinic Gd$_2$O$_3$}, {\em J. Am. Ceram. Soc.\/}, {\bf 56}(6), 327--330
  (1973).

\bibitem{Hunter_hfo2}
S.~L. Dole, O.~Hunter, and F.~W. Calderwood, \enquote{Elastic Properties of
  Stabilized HfO$_2$ Compositions}, {\em J. Am. Ceram. Soc.\/}, {\bf 63}(3-4),
  136--139 (1980).

\bibitem{Porter_spinel}
D.~F. Porter, J.~S. Reed, and D.~Lewis, \enquote{Elastic Moduli of Refractory
  Spinels}, {\em J. Am. Ceram. Soc.\/}, {\bf 60}(7-8), 345--349 (1977).

\end{thebibliography}
\end{document}